\documentclass[a4paper]{jpconf}
\usepackage{graphicx}
\begin{document}

\title{A dynamical description of neutron star crusts}

\author{V. de la Mota, F. S\'ebille and Ph. Eudes}

\address{SUBATECH, Ecole des Mines, Universit\'e de Nantes, CNRS/IN2P3, Nantes, France}

\ead{delamota@subatech.in2p3.fr}

~\\[2ex]
\small
\noindent Comments: Talk given by Virginia de la Mota at the 11th International Conference on Nucleus-Nucleus Collisions (NN2012), San Antonio, Texas, USA, May 27-June 1, 2012. To appear in the NN2012 Proceedings in Journal of Physics: Conference Series (JPCS)\\
Subjects: Nuclear Theory (nucl-th); Solar and Stellar Astrophysics (astro-ph.SR) 
\normalsize

\begin{abstract}
Neutron Stars are natural laboratories where fundamental properties of matter under extreme conditions can be explored.  Modern nuclear physics input as well as many-body theories are valuable tools which may allow us to improve our understanding of the physics of those compact objects.

In this work  the occurrence of exotic structures in the outermost layers of neutron stars is investigated within the framework of a microscopic model. In this approach the nucleonic dynamics is described by a time-dependent mean field approach at around  zero temperature. Starting from an initial crystalline lattice of nuclei at subnuclear densities the system evolves toward a manifold of self-organized structures with different shapes and similar energies. These structures are studied in terms of a phase diagram in density and the corresponding sensitivity to the isospin-dependent part of the equation of state and to the isotopic composition is investigated. 
\end{abstract}

\section{Introduction}

Neutron stars (NS) are known to be born after the core collapse of supernovae explosions. After its birth the NS has cooled below 10$^9$ K and a solid crust is supposed to be formed in the outermost layers. Even if the crust contains only a small part of the star, it plays an important role in its evolution. Indeed, the properties of the crust, notably the transport properties, can strongly influence processes like cooling, accretion and X-ray bursts.\\
Heavy ions and the crust of  NS  share basic many-body aspects due to the fact that they are made of interacting nucleons. Having more neutrons than protons, many physical properties of both kind of objects are  related with the asymmetry dependence of the nuclear interaction.\\

In this work we were interested in dynamical aspects of NS crusts, in particular, with the occurrence of exotic structures. In that region, at densities  just below the saturation value, the nuclear matter is said to be ``frustrated'': subjected to the competition of the short-range nuclear attraction and the long-range Coulomb repulsion. As a consequence of this competition nuclei may organize forming non-spherical composites. Structures such as rod-like and slab-like nuclei, the so called nuclear ``pasta'', were initially predicted to exist in the ground state of nuclear matter at subnuclear densities by static approaches  \cite{pasta0}.  Assuming well defined symmetries, these models suggested the appearance of these structures in a well defined order from spherical to cylindrical, planar, complex structures and finally to uniform matter, with increasing density. This typical sorting of structures was confirmed by recent works \cite{pasta1} and suggested that pasta phases can be formed in the cooling process of hot NS and during the compression stages taking place during the collapse of supernova cores \cite{watanabe12}. In both cases it is worthwhile to give a dynamical description of pasta formation.

The interest in structural phases in NS crusts is double. On one side, their existence may have various astrophysical consequences. Their presence can influence the scattering of neutrinos and modify the neutrino trapping with the corresponding consequences on the processes of supernova core collapse or NS cooling. Another aspect which can be sensitive to pasta formation are the elastic properties of the star (namely those phenomena depending on the crust-core boundary properties) which may influence gravitational wave radiation. On the other side, from the nuclear physics point of view, their formation can be connected with fragment formation in heavy ion collisions. These phenomena share many aspects  since they are both driven by volume, Coulomb and surface energies. In particular, the sensitivity to the nuclear equation of  state has been suggested \cite{horowitz04} as in cluster formation. 

In this work we present a model describing the dynamics of nuclear matter in the crust of a NS.  It is an extension of a dynamical approach initially developed in order to describe heavy ion collisions at intermediate energies \cite{dywan}.
In contrast to the many previous studies employing static frameworks, this model allows to
simulate the dynamical processes in inhomogeneous nuclear matter using a large number of nucleons without any assumptions on the structure of nuclear matter. Since these structures involve a large number of low energy configurations, in this work a pure mean field description of the nuclear dynamics has been performed. In this approach, starting from initial crystalline lattices of nuclei, with different symmetries, non spherical structures occur as the result of microscopic self-organization processes.  The survival of those meta-stable equilibrium structures has been checked over thousands fm/c \cite{sebille09}. In this work, the influence of the equation of state (EOS), of the isotopic compositions and of the lattice perturbations on the formation of exotic structures are sudied.

This contribution is organized as follows. In Section 2 the bases of the model are presented. In Section 3 we analyze the occurence of structures in  Oxygen lattices with different isotopic composition. The effects of the EOS and the response to lattice perturbation are discussed. In Section 4 our concluding remarks are given.

\section{A dynamical model for the neutron star crust}

At temperatures $T\ll$ 1 MeV and densities between 10$^6$g/cm$^3$ at the surface to 3$\times 10^{13}$g/cm$^3$ at the interface with the core, the stellar matter can be modelled by a neutral  mixture of nuclei, electrons and free neutrons. Nuclei are expected to form a crystalline lattice immersed in a degenerate relativistic gas of electrons, which may be described by a uniform density 
distribution \cite{crystal}. In order to describe the microscopic dynamics of the nuclear many-body system we have developed a dynamical approach, which is based on the DYWAN model \cite{dywan} for heavy-ion collisions at intermediate energies. This approach essentially gives a prescription for solving an extended time-dependent Hartree-Fock (ETDHF) equation for the one-body density matrix,
of the form:
 \begin{equation}
\label{etdhf}
 i\hbar \dot{\rho}=[{\bf h}(\rho),\rho]+ \mathcal{K}(\rho). 
\end{equation}
where $\rho$ is the one-body density matrix, ${\bf h}(\rho)$ is the one-body Hamiltonian and $ \mathcal{K}(\rho)$ is the collision term, providing the irreversible evolution of the system due to the coupling of $\rho$ with higher order multiparticle correlations. In the present calculations we are essentially concerned with low energy configurations of the system. We will suppose that transitions towards excited single particle states are inhibited and that the dynamics is mainly governed by the mean field.

Equation (\ref{etdhf}), with $\mathcal{K}(\rho)$=0, is solved by projection of the one body space of states onto a convenient basis.These functions are splines \cite{splines}, a particular case of  wavelets, which are mathematical (non analytical) objects with many interesting properties and which behave like a moving basis. 

In a first step, a static Hartree-Fock (HF) self-consistent procedure is 
implemented in order to get nuclear composites either in their ground states 
or in excited states according to mechanical or thermal constraints.
With these nuclei, prepared in a single Wigner-Seitz (WS) cell,  a super-cell is constructed
by stacking unitary cells in three dimensions. Their number and symmetries are different depending on the kind of lattice we are interested in. The complete three-dimensional lattice is built up from the super-cell under periodic boundary conditions. In this work  simple cubic cells (SCC) have been
considered, but other symmetries can be arbitrarily chosen.

For this calculation we have chosen, for simplicity, a density-dependent zero-range effective
interaction, with the following self-consistent field:
\begin{eqnarray}\label{v}
V^{HF}_q(\rho,\xi)&=&\frac{t'_0}{\rho_\infty}~\rho~+~\frac{t'_3}
                    {\rho_\infty^{\nu+1}}~
                    \rho^{\nu+1}~
                  +~\frac{c}{{\rho_{\infty}}^2}
                 \xi^2~+~\frac{4qc}{{\rho_{\infty}}^2}\rho \xi+\\
\nonumber		 
		 & & \frac{\Omega}{3{\rho_{\infty}}^2}\xi^2
		 +\frac{4q\Omega}{3{\rho_{\infty}}^2}(\rho-\rho_{\infty})\xi
		 +V_q^C ,
\end{eqnarray}
where $\rho_n$ and $\rho_p$ stand for neutron and proton densities,
$\rho=\rho _n~+~\rho _p$, $\xi=\rho _n~-~\rho _p$,
$q$=1/2 for neutrons and -1/2 for protons, $\rho_{\infty}$=0.145 fm$^{-3}$ 
is the saturation density of infinite nuclear matter and $V_q^C$ is the Coulomb 
potential. The calculation of $V_q^C$ is performed using the Ewald 
summation techniques \cite{ewald}, which is adapted to the calculation of long range potentials 
in periodic systems. It consists basically in recasting the Coulomb potential into two convenient 
terms, each one of them  can be calculated in a fast and efficient way \cite{sebille09}.

The parameters of the force are related with those corresponding to the usual 
Skyrme force, $t_0$, $t_3$, $x_0$ and $x_3$, by the following relationships:
\[t'_0=\frac{3\rho_\infty}{4}t_0~~~~~~
t'_3=\frac{3(\nu+2)\rho^{\nu+1}_\infty}{48}t_3~ ,\]  
with $x_0=x_3=-1/2$, $\nu$=1/6,  and:
\[t'_0/\rho_{\infty}=-356\mathrm{MeV~{fm}}^3~~~~~~ 
t'_3/\rho_{\infty}^{\nu+1}=303 \mathrm{MeV~{fm}}^{3(\nu+1)} ~.\]

\noindent The parameters $c$ and $\Omega$ are related to the coefficients of the mass formula $J$ and $L$ corresponding to, respectively, the volume-symmetry and to the density-dependent symmetry energy as follows:
\[c=J-\frac{h^2}{6M}k_F^2
\]
\[L= 6c+\Omega+ L_{kin}
\]
$L_{kin}$ is the kinetic energy contribution to $L$.
These parameters are still uncertain because they are not completely constrained from nuclear data, then a study of their influence on the overall dynamics seems necessary. In this work $c$ has been 
fixed to 20 MeV and $\Omega$ varied in the range -125 and -75 MeV in order to analyze the sensitivity  of the dynamics to this interaction. The density dependence of the symmetry energy has been determined here in a phenomenological way according to its current estimate at the saturation value in pure neutron matter. The values of the parameters reproduce the principal
static characteristics of nuclei, as binding energies, radii and equilibrium densities. The associated incompressibility modulus in symmetric matter $K_{\infty}$= 200 MeV corresponds to a ``soft'' EOS.

\begin{figure}[h]
\begin{center}
\includegraphics[width=8cm,height=6.5cm]{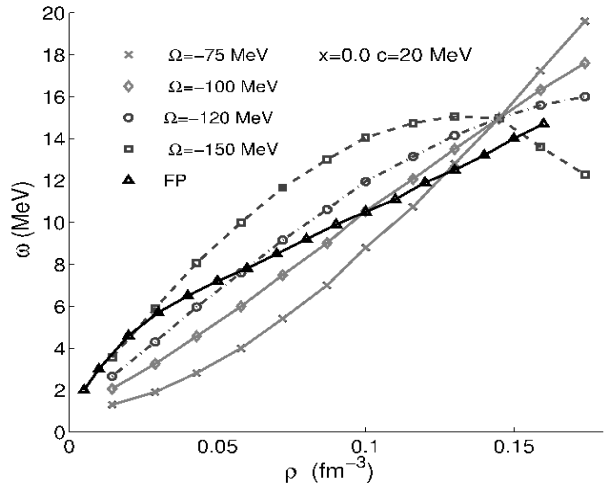}
\end{center}
\caption{\label{eos} Neutron energy density per baryon as a function of the
density  at different values of the asymmetry parameter $\Omega$. The results of
Ref. \cite{panda} are in triangles.}
\end{figure}

The energy density per baryon $\omega$ is defined as:
\[
\omega=\frac{\varepsilon}{\rho}
=\frac{\int V^{HF}_q~ d\rho}{\rho} +\omega_{kin}, 
\]
where $\omega_{kin}$ corresponds to the kinetic contribution. In Fig. \ref{eos} the calculated
values of $\omega$ in pure neutron matter are represented as a function of the mean density for different values of the asymmetry parameter $\Omega$. Together with our calculations, the results of Ref. \cite{panda}, in the framework of a microscopic variational model,  are plotted in triangles.

\begin{figure}[h]
\begin{center}
\includegraphics[width=9cm,height=7cm]{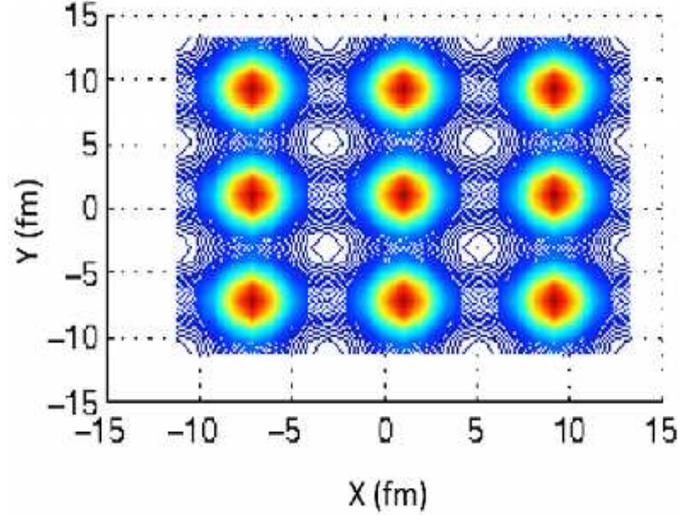}
\end{center}
\caption{\label{crystal}  Density profiles of $^{16}O$ lattice on the (X,Y) plane.  }
\end{figure}

Following the iterative procedure,  Oxygen nuclei have been prepared in WS cells with the force given by Eq. (\ref{v}). The three-dimensional lattice is then constructed with replicas of basic SCC with periodic boudary conditions. In this case the supercell is composed by 27 WS cells. In Fig. \ref{crystal} is 
represented the projected density profiles on the symmetry plane (X,Y) of a SCC lattice of Oxygen isotopes with proton fraction $x$=0.5 and mean density $\langle \rho \rangle$=0.02 fm$^{-3}$. Other kind of lattices with different symmetries and species have been considered in Ref. \cite{sebille11}.

As stated before, once the initial conditions are given, the system is led to evolve according to 
Eq. (\ref{etdhf}) turning off the collision term. The resulting equation is solved by projection of single particle wave functions onto a spline wavelet basis  $\alpha (\vec r, t)$ \cite{splines}. These functions are orthogonal, they present defined symmetries and compact support and, even if they are not analytical they can be  approximated by simple analytical forms. At the same time, they provide strongly compressed representations of the system with an accurate description of single particle  wave functions. 

In configuration space wavelets are functions of a set of correlated generalized coordinates 
$\lbrace \vec \xi,\vec \chi, \vec \pi, \vec \phi\rbrace$, 
which correspond to first and second moments. In the $x$ dimension they read :
\[<x>=\xi_x ~~~ <(x-\xi_x)^2>=\chi_x ~~~<p_x>=\pi_x ~~~
<(p_x-\pi_x)^2>=\phi_x ~,\]
with similar expressions for $y$ and $z$ components.

It can be shown \cite{sebille09} that wavelets satisfy modified TDHF-like equations, while  their corresponding occupation numbers evolve through a master equation. The equations of motion for the generalized coordinates can be obtained from a viariational principle \cite{sebille09}:
\begin{eqnarray}
\label{dotq1}
\dot{\xi}&=&\frac{\pi}{m} + \frac{\partial}{\partial \pi} {\cal{V}}\\
\label{dotq2}
\dot{\pi}&=&-\frac{\partial}{\partial \xi} {\cal{V}}\\
\label{dotq3}
\dot{\chi}&=&\frac{4 \gamma \chi}{m} - \frac{\partial {\cal{V}}}{\partial \gamma}\\
\label{dotq4}   
\dot{\gamma}&=&\frac{\hbar^2} {8 m \chi^2} - \frac{2 \gamma^2} {m}
    -\frac{\partial}{\partial \chi} {\cal{V}} ~. 
\end{eqnarray}
${\cal{V}} $ being the wavelet transform of the effective nuclear potential 
${\cal{V}}=<\alpha|V^{HF}|\alpha>$. It is important to underline that the 
self-consistent mean-field is a function of the density extended to the
overall supercell and not only to a single WS cell.

Equations \ref{dotq1}-\ref{dotq4}, which solve the TDHF equation for the nucleonic system, are Hamilton-like equations for the centro\"ids and the widths of the moving basis of wavelets.

\section{From crystalline to exotic structures}

\begin{figure}[h]
\begin{center}
\includegraphics[width=12cm]{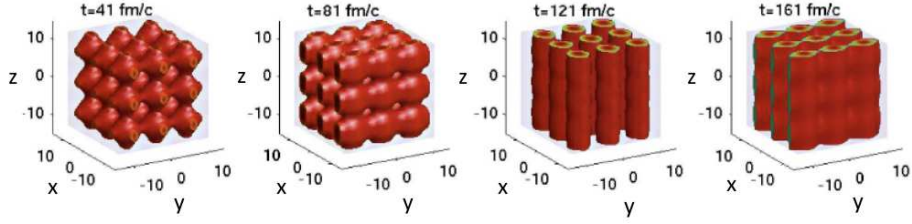}
\end{center}
\caption{\label{3d_evolution} Time evolution of neutron density in oxygen SCC lattice with $x_p$=0.1, $\langle \rho \rangle$=0.072 fm$^{-3}$ and  threshold density 0.065 fm$^{-3}$.}
\end{figure}

Let us now consider the initial arrangement of Oxygen isotopes, slightly deformed along the z axis, in a SCC lattice with proton fraction $x$=  0.1 and  mean density $\langle \rho \rangle$=0.072 fm$^{-3}$. Some snapshots of the evolution are shown in Fig. \ref{3d_evolution}. 
The system goes through some preferred structures, in this particular case, it oscillates between rods and slab-like shapes. These plots represent points in configuration space for which the density is higher than a reference value: the threshold density $\rho_t$. The corresponding value in Fig. \ref{3d_evolution} is 0.065 fm$^{-3}$. For a given configuration the choice of this quantity determines the underlying structures. In consequence the occurrence of a given shape depends not only on the mean density but also on the choice of $\rho_t$. Even more, for a given choice of both quantities, different embedded structures can be found.

In order to characterize unambiguously these structures we have utilized the principles of morphology recognition from Morphological Image Analysis (MIA) techniques, which assign numbers to the different shapes according to some geometrical prescription \cite{mia}.

\begin{center}
  \begin{figure}[htbp]
  \includegraphics[width=8cm]{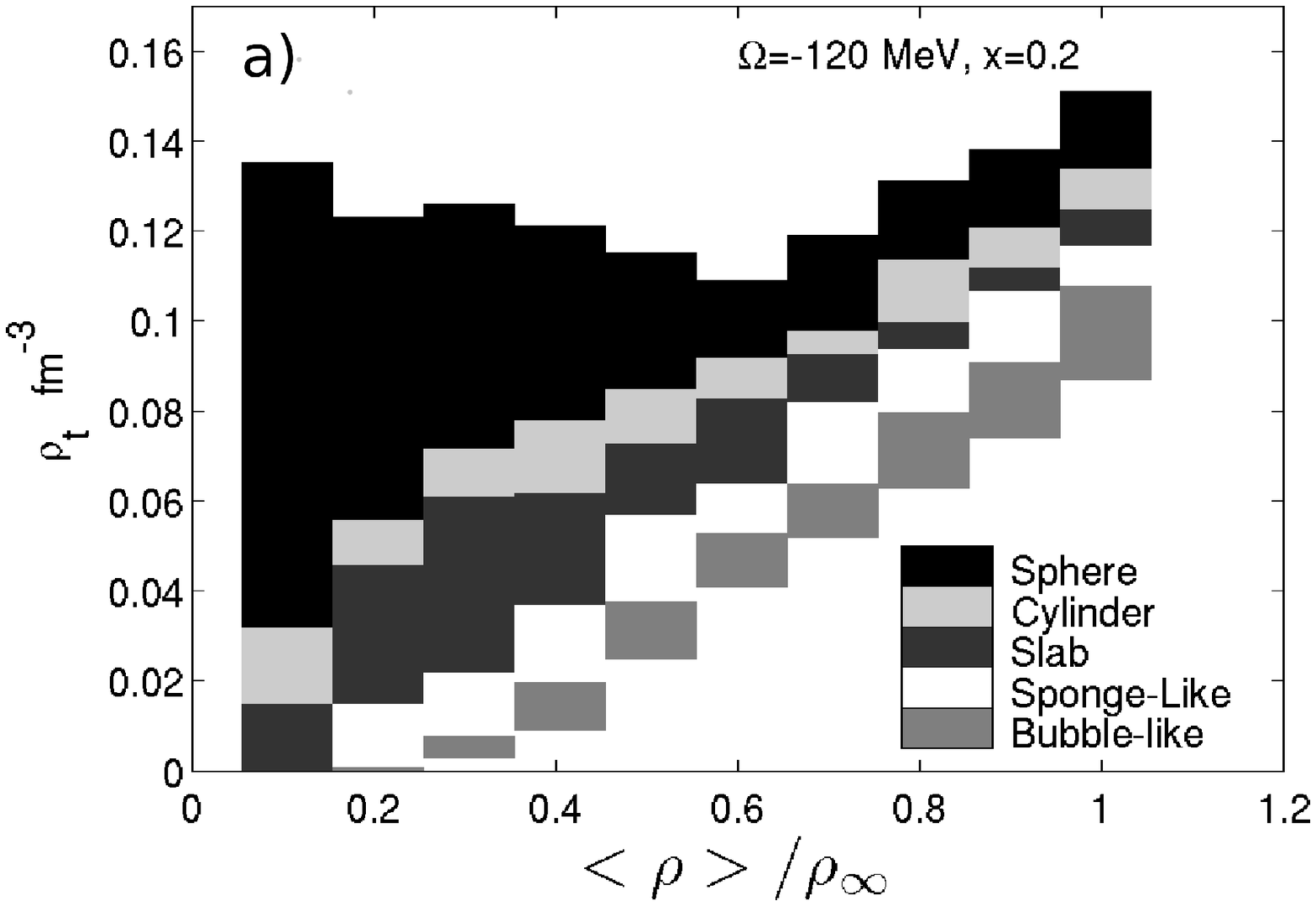}
  \includegraphics[width=8cm]{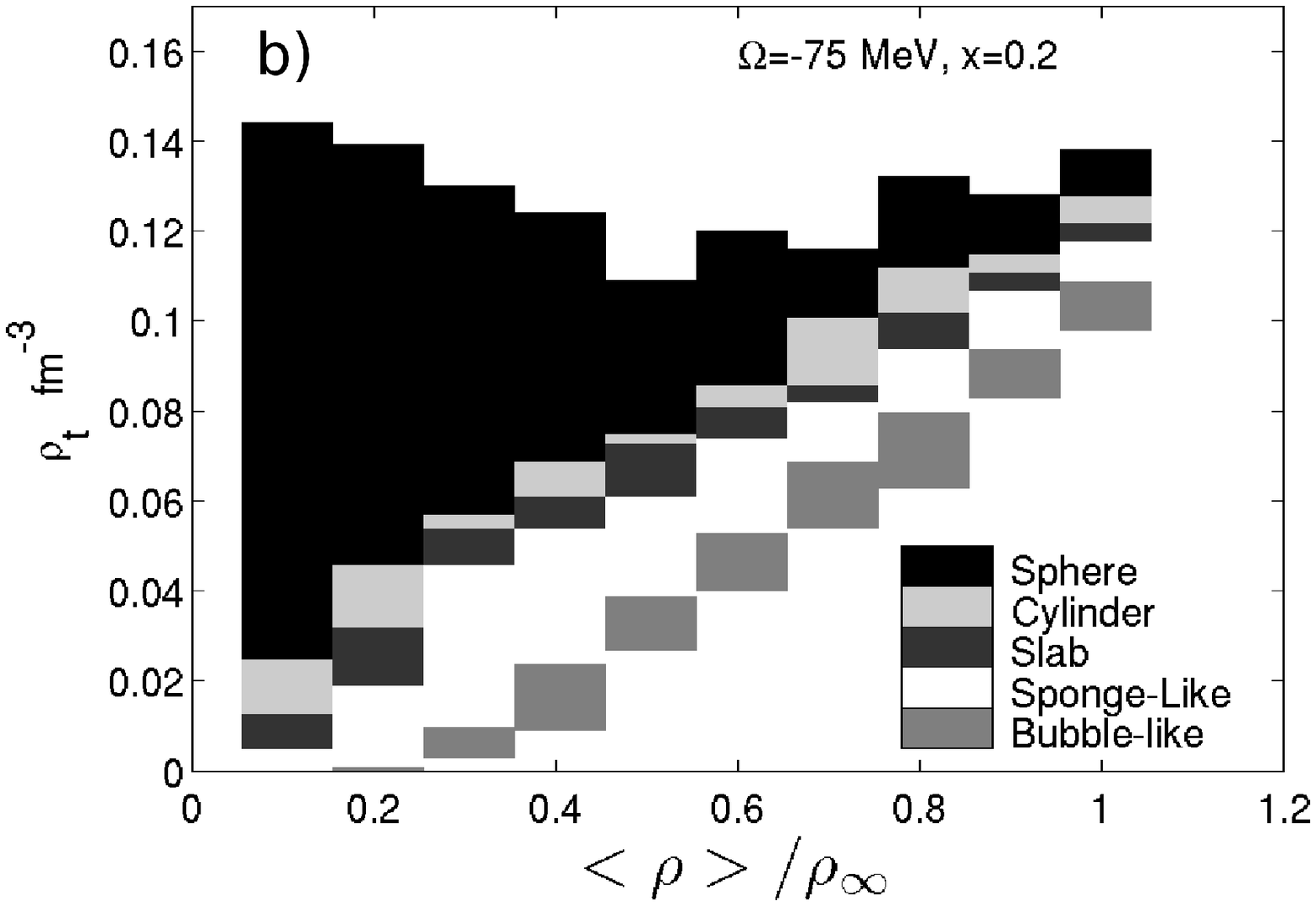}
  \caption{ Neutron threshold density versus the
   neutron mean density normalized to the saturation value for two
   values of $\Omega$ and for proton fractions $x$=0.2. 
  }\label{diagram}
  \end{figure}
  \end{center}

Plotting the different structures in the density plane ($\rho_t$ vs $\langle \rho \rangle$) gives an overall view of the most probable structures occurrence for different initial configurations. In Fig \ref{diagram} the corresponding diagrams for Oxygen isotopes SCC lattices is represented in the asymmetric $x$=0.2 case, for two values of the asymmetry parameter $\Omega$: -120 MeV (a) and -75 MeV (b). The different regions in grey scale correspond to the various structures, from black colour for spherical nuclei to white colour for sponge-like shapes.
In any case, for a given threshold density, the five standard types of pasta phases emerge naturally and  in the same order with increasing density, as predicted by static models \cite{pasta0} and confirmed by recent calculations \cite{pasta1}. 
In this example the system experiences some sensitivity to the asymmetry-dependent contribution to the effective force. Indeed, in Fig. \ref{diagram} the stiffer force, corresponding to $\Omega$=-75 MeV, is shown to favour the occurrence of spherical and sponge-like structures, while restraining the appearance of cylinder and slabs.  This is consistent with the fact that in the stiff case the potential is more repulsive, namely at high mean densities. The consequence is that more and more neutrons are dripped or localized at the surface. In the last case, quasi-free neutrons contribute either to build heavy spherical nuclei or to link neighbouring clusters is all directions, giving rise to the so-called sponge-like structures.  

\section{Effects of lattice perturbations}
\begin{figure}[htbp]
\begin{center}  
  \includegraphics[width=8cm]{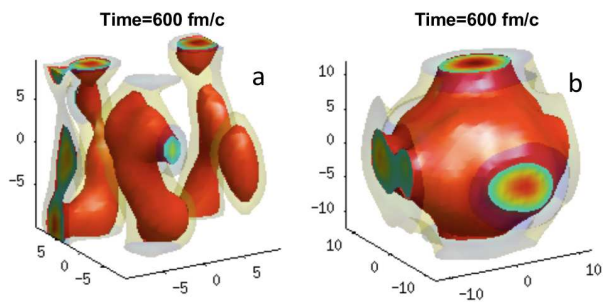}
   \caption{ Snapshots of neutron density at t=600 fm/c in perturbed oxygen SCC lattice with $\langle \rho \rangle$=0.06 fm$^{-3}$  for (a) $x_p$=0.5 and (b) $x_p$=0.2. 
  }\label{perturb}
 \end{center} 
 \end{figure}

As underlined by other authors \cite{magierski02} the inner crust of a neutron star is an extremely complex system in which the appearance of disordered phases can be favored. There is a special interest to investigate how nuclear matter, for a given proton fraction, self-organizes dynamically with the introduction of slight perturbations of the initial conditions.
In order to study the feasibility of those disordered phases, we considered SCC lattices of oxygen isotopes, with $x_p$=0.2 and 0.5, the positions of which have been initially slightly shifted 
at random from the lattice sites. In Fig. \ref{perturb} the snapshots of the  supercell density at t=600 fm/c is represented in the cases in which the proton fractions are $x_p$=0.5 (a) and 0.2 (b). In both cases the average density is $<\rho>=0.4 \rho_{\infty}$,  the threshold densities $\rho_{t}$ are 0.04  and 0.05, respectively.\\
Starting from similar initial conditions the system evolves towards unlike configurations depending on their corresponding proton fraction. In the symmetric case the system organizes in several intermediate mass clusters, while in the asymmetric case it evolves towards a unique massive cluster occupying the entire supercell.  In any case, a gradual loss of the initial symmetries is observed in both pictures. From an ordered array of regular deformed nuclei the system goes through a heterogeneous configuration of clusters with different masses and shapes.
It is worthwhile to underline that in Ref. \cite{sebille09} the model has been shown to be extremely well conditioned, conserving metastable configurations during time intervals of several thousands of fm/c. This fact permitted to check the corresponding numerical accuracy and reject possible numerical instabilities. 

The small initial perturbations then generate sufficiently strong density fluctuations so as to provoke a different rearrangement of matter molding heavy aggregates. The fact that at lower proton fraction a single (infinite) aggregate occurs can be interpreted as a consequence of the increase of neutron wave functions diffusivity. This is a direct  consequence of wavelet spreading which results from the dynamical evolution. 
Indeed, as the proton fraction decreases, the number of  dripped neutron increases and nucleons become more delocalized with significant overlap in configuration space.  The competition between nuclear and Coulomb interactions, which is local, covers broader spatial regions in this case.

\section{Conclusions}

In this work we present a non dissipative, self-consistent, dynamical description of nuclear matter in conditions of density and temperature similar to those which should exist in the outermost layers of neutron stars. \\
In this description the small excitation energy which is initially deposited in deformation allows the system to explore a landscape of structures. The occurrence of most probable shapes is analyzed by means of structure diagrams in the threshold density versus mean density plane. Not only the ordered five standard types of pasta emerge naturally but also other intermediate shapes between them which can be characterized through morphological analysis techniques.\\
Phase diagrams are shown to be sensitive to a variation in the intensity of the isovector part of the 
self-consistent potential. The observed trend is the increase of spherical or sponge-like structures with the stiffness of the force.\\
In the present pure mean-field description lattice symmetries  are preserved along the temporal evolution of the system. In all cases, the effects of perturbing the initial lattice is to break  those symmetries. Nevertheless the characteristics of the final distribution depend on the isotopic composition. Symmetric systems organize in several intermediate mass structures, while asymmetric ones collapse in a unique massive cluster occupying the entire supercell.

In this work a simple phenomenological local potential has been implemented.
The formation and transitions between the observed non-equilibrium structures can be  strongly 
modified using more complex non-local effective forces  and beyond a pure mean-field description. The study of the transport properties of the crust can be then performed in more reliable conditions. These investigations are currently in progress.

\end{document}